# The Internet of Value: Integrating Blockchain and Lightning Network Micropayments for Knowledge Markets

Ellis Solaiman[1] and Jorge Robins[1]

[1]*School of Computing, Newcastle University, Newcastle Upon Tyne, UK.

E-mail(s): ellis.solaiman@ncl.ac.uk; jrobins25@hotmail.com

**Abstract**

Q&A websites rely on user-generated responses, with incentives such as reputation scores or monetary rewards often being offered. While some users may find it intrinsically rewarding to assist others, studies indicate that the promise of payment can improve the quality and speed of answers. However, traditional payment processors often impose minimum thresholds that many Q&A inquiries fall below. The introduction of Bitcoin established a basis for direct digital value transfer, yet the practical application of frequent micropayments has remained challenging. Recent advancements like the Lightning Network now enable frictionless micropayments by reducing transaction costs and minimising reliance on third-party intermediaries. This development fosters the realisation of an "Internet of Value", where transferring even very small amounts of money is as straightforward as sharing data. This study investigates the integration of Lightning Network-based micropayment strategies into Q&A platforms, aiming to establish a knowledge market free from traditional minimum payment barriers. A questionnaire survey was conducted to address the gap below the $2 payment level identified in previous research. The responses confirmed that incentives for asking and answering questions weaken as the monetary amount decreases. The findings reveal that even minimal payments, such as £0.01, serve as significant incentives, encouraging users to enhance the quality and effort of their responses. The study recommends adopting micropayment incentive strategies for service-oriented applications, particularly the Q&A use case. By leveraging the Lightning Network to remove entry barriers, a more open marketplace can emerge, potentially improving user engagement and outcomes. Further research is needed to determine if users will follow through on their reported intentions when actively spending their funds.

**Keywords:** Micropayments, Lightning Network, Blockchain, Q&A Websites, Incentives

# 1 Introduction

The Internet of Value (IoV) represents a transformative concept wherein valuable assets, ranging from digital currencies and securities to intellectual property and real estate, can be exchanged directly between parties. This direct exchange is facilitated without the traditional reliance on intermediaries such as banks, brokers, or escrow services. The primary enabler of IoV is the integration of blockchain technology, along with off-chain scaling solutions like the Lightning Network (LN), which ensures secure, transparent, tamper-proof, and efficient transactions [1]. The IoV operates similarly to how data is shared online, enabling direct asset exchanges without intermediaries. Yet, similar to data sharing, IoV contends with privacy challenges. In" Life after Google," Gilder highlights tensions between data exchange and privacy. The dominant internet model, led by large technology firms, centres on data collection and monetisation, thus diminishing user privacy [2].

Individuals have grown more conscious of the personal information collected about them. Advertising has become pervasive, and many users have voiced their dissatisfaction with frequent ads and their impact on the internet experience. For instance, a study by [3] discovered that by February 2016, ad blockers were installed by up to 37% of users in Germany, 18% in the US, and 17% in the UK. They studied more than 2 trillion web interactions involving 2 million users. A vital alternative to data and information exchange using search engines is the concept of the knowledge market. Knowledge markets facilitate the creation and exchange of intellectual capital. Their efficacy rests on dynamic interactions of knowledge creation and transfer [4]. Moreover, plummeting costs for computing devices and internet connectivity have fostered better infrastructures for knowledge exchange [5]. In his seminal text, Mises defined human action as purposeful behaviour [6]. From this viewpoint, reason is deployed to optimise economic objectives. In a knowledge market, an answer provider requires a tangible incentive to participate.

This report explores treating knowledge as a tradable good, rather than a public resource freely disseminated. Historically, micropayments were impeded by the cost overheads of financial authentication and adjudication [7]. Reversal options also undermined accountability, as even small transactions could be rescinded long after completion. Bitcoin introduced a decentralised time-stamping and value transfer system, thus reducing reliance on intermediaries [7]. Building on earlier proposals [8], Bitcoin employed a distributed ledger architecture. Transactions are recorded in blocks interconnected by a Merkle root hash [9]. Altering past transactions becomes virtually impossible without detection. Even so, frequent, ultra-low-value micropayments continued to face hurdles due to transaction fees and throughput constraints on the Bitcoin blockchain. This is where the Lightning Network becomes important. As a layer-2 protocol atop Bitcoin, LN diminishes fees and confirmation delays, unlocking viable micropayment channels [10]. As of August 2024, LN transactions incurred remarkably low fees, making it feasible to send fractions of a penny [11]. This innovation enables everyday internet tasks, from reading articles to using search engines, to be monetised without a reliance on intrusive data collection or unwieldy fees. By leveraging LN-based micropayments, users can maintain increased control of their data and privacy, while still accessing a wealth of online content. Micropayments, therefore, offer fresh prospects. Consumers unwilling to be subjected to advertisements or data mining can directly pay for digital content or services. While this project concentrates on a consumer-to-consumer (C2C) scenario, LN's channels adapt readily to both business-to-consumer (B2C) and business-to-business (B2B) payments. Embracing LN may transform online economic interactions on a grand scale, creating diverse and flexible usage models. If LN-enabled micropayments reduce unanswered questions and boost answer quality and speed in Q&A sites, this approach can enhance user satisfaction. New revenue models could emerge for providers. For instance, knowledge markets might charge tiny fees for user actions, passing a portion back to content creators. Over time, these small incentives could accumulate, especially for users in developing regions, potentially forming part of one's livelihood.

This report contributes a proposed integration of LN micropayments with blockchain-based knowledge markets. We define micropayments here as any sum below £1.00, but the demonstration extends further, highlighting transactions below £0.01. Although more involved questions could command higher bounties, LN micropayments widen market accessibility, especially for individuals priced out by conventional processors' thresholds. A usability evaluation will assess LN's competitiveness against existing solutions. Additionally, a questionnaire survey will explore whether micropayments, on a scale significantly smaller than previous studies, still foster robust incentives. Finally, we will examine if users' self-reported motivation remains when they must spend their own funds.

## 1.1  Outline of the Paper

To contextualise our discussion, the paper is organised as follows. Section 2 explores smart contracts and the nascent research on using blockchain in knowledge-sharing contexts. Section 2 also reviews relevant literature about Q&A economic models and micropayment viability. Section 3 outlines the research methodology, including the design of our questionnaire instrument. The subsequent subsections detail the questionnaire structure and results. Our conclusions in Section 4 highlight promising insights and future research directions, including areas like sustainability, live prototyping, and longitudinal analysis.

## 2  Background

### 2.1  Smart Contracts

In modern society, contracts are ubiquitous, ranging from informal verbal agreements to formally documented and legally binding ones. Their prevalence is indisputable, with origins tracing back to ancient civilisations [12]. The proliferation of electronic commerce has led to an exponential increase in the number of contracts a business can engage in, posing challenges in managing the complexities of digital markets.

Creating a contract involves substantial investments of time, money, and human resources. This process encompasses hiring specialised personnel to draft and validate contract terms, negotiating between parties, and resolving disputes with third-party mediators. Szabo [13] introduced the concept of smart contracts as self-executing digital agreements that incorporate the conditions for their execution. In this study, we use the term "smart contract" to refer to digitally encoded contracts that function either with or without a blockchain system. The objective of smart contracting is to streamline the process of contract creation and execution while minimising costs. To achieve this goal, a smart contract must be formulated in a programming language capable of expressing all contractual details while eliminating ambiguities [14–18].

For a contract to be suitable for monitoring and enforcement by any contract compliance system, it must be formally expressed in a language capable of representing contractual requirements, including legal stipulations, clauses, internal policies, involved parties, and their actions. The objective of an effective contract monitoring and compliance solution is the automation and execution of contracts with minimal human intervention. Consequently, the language employed to define the contract must be unambiguous and exact, eliminating the need for manual conflict resolution.

The surge in smart contract research can be attributed to the emergence of blockchain technology, marked by the creation of Bitcoin, Ethereum, Hyperledger, among others. Smart contracts have proven useful across various domains, such as education [19–22], Internet of Things [23, 24], and Service Level Agreements (SLAs) [25–29]. Notably, research on multi-blockchain interoperability has illustrated how verifiable and decentralised querying frameworks can be applied in heterogeneous blockchain environments to improve trust and transparency [30, 31]. Beyond single-chain implementations, advanced architectures for blockchain-based provenance tracking and sustainable data management have also been explored [32, 33]. Indeed, the concept of smart contracts predates blockchain technology, allowing for implementations in centralised, distributed, and hybrid architectures [34–36].

Research in smart contracts and blockchain spans multiple domains, including performance evaluation [37] and simulation studies [38–40]. Work has also emerged on using blockchain as a foundation for federated learning security [41] and on integrating IoT monitoring with smart contracts derived from SLAs [42, 43]. In particular, a key theme in these studies is data integrity and provenance, which is essential for guaranteeing trust in diverse ecosystems, from supply chains to sensor-based environments [31–33, 44].

In the context of Q&A websites, integrating smart contracts can facilitate micropayments for user-generated responses. While early blockchain solutions laid the groundwork, the high costs and latency associated with on-chain transactions posed challenges for true micropayment incentives. The introduction of the Lightning Network has addressed these issues by allowing near-instant, low-cost off-chain transfers. This advancement realises an "Internet of Value" where transferring funds, even very small amounts, is as straightforward as sharing data [45]. Consequently, it becomes feasible to reward minimal contributions, encouraging users to offer more effort and expertise without being constrained by traditional payment thresholds.

By employing Lightning Network-based micropayment strategies, Q&A platforms can operate as knowledge markets that remove minimum payment thresholds. This supports a free-market approach, potentially improving answer quality and response times. Nonetheless, further research is needed to ascertain whether users will indeed follow through on their reported willingness to spend funds when actively engaging in LN-driven micropayment ecosystems.

## 2.2 Literature Review

Economists generally associate higher prices with greater willingness to supply, and thus higher monetary incentives might be expected to increase the satisfactory answer rate in a knowledge market. Google Answers (GA) was a monetised knowledge market that operated from 2002 to 2006 and drew research attention. Users could pay between $2 and $200 for a question and could also tip answerers. Studies by [46] showed that paying more for an answer in GA produced higher-quality answers, longer answers, and higher effort from answerers. However, a field study by [47] contradicted this view, finding that paying more only resulted in longer answers, not better quality answers. Both studies agreed that price and tips significantly influenced answerers' efforts. The importance of reputation systems was also highlighted, with higher-reputation answerers providing significantly better answers, as discovered by [47].

Psychology supports this view, with [48] arguing that powerful nonpecuniary motives shape human behaviour. They warn that focusing only on contract theory and principal-agent theory in economics might restrict attention to risk avoidance and income through effort, neglecting "the desire to reciprocate, the desire to gain social approval, and the intrinsic enjoyment arising from working on interesting tasks." However, [49] suggests that higher-quality answers from higher-reputation users could be due to their having learned to provide answers that questioners value most as they gained experience. In industry, client support models follow SLA metrics such as "Time to Response" and "Time to Resolution," but this is not well-studied in community-driven or paid-for Q&A sites.

[50] argue that improving not just the quality but also the speed of acceptable answers can enhance user experience and engagement. They found that incentive systems do not recognise non-frequent answerers who often tackle more challenging questions. On average, frequent answerers respond within 81 hours, while non-frequent answerers take 316.4 hours. The authors suggest offering greater rewards for more demanding questions to avoid leaving them unanswered.

The literature examining the efficacy of micropayments as an incentive mechanism in markets is limited. While Bitcoin introduced the concept of frictionless digital payments, transaction fees and speed constraints initially limited the scope for true micropayments. The Lightning Network now addresses these issues by enabling scalable off-chain transactions, making it more practical to test micropayments as an incentive model at scale.

A study by [51] investigated incentive mechanisms in participatory sensing tasks and found that users preferred tangible and predictable rewards over uncertainty. Specifically, 62.5% of participants chose a fixed micropayment scheme compared to only 4.16% for variable micropayment and lottery-style schemes. Researchers in [52] examined compensation in studies where participants wore sensors over extended periods and found that 53% of participants self-reported a preference for a $50 bulk payment. However, their observations indicated that smaller, continuous micropayments increased user engagement and attentiveness. This suggests that even if users claim to prefer lump sum payments, micropayments may yield better compliance and retention. Normative bias may explain the discrepancy between reported and observed behaviour. Authors of [52] argued that micropayments faced resistance from governments and sellers, making them unsuitable for widespread adoption. However, the emergence of LN-based micropayments challenges this view, showing that technical solutions can reduce barriers and create new economic models. Bitcoin SV (BSV), for instance, follows the original Bitcoin protocol and

is Turing complete, allowing for arbitrarily complex logic to be programmed into transaction scripts while also facilitating micropayments.

The debate on micropayments remains open, with empirical support and opposition to their success. If price and tip incentive models can produce fewer unanswered questions, higher quality answers, and improved response/resolution times, then LN-enabled micropayment incentive models may be well suited to knowledge markets. This is especially true if a significant share of questions is valued below £1.

Given the scarcity of research in this area, this paper explores the synthesis of these ideas to gain greater insight into how users of knowledge markets perceive and respond to LN-driven micropayment incentives. This approach may serve as an alternative to reputation-based systems, potentially reshaping how knowledge is exchanged and valued online.

## 3  Methodology

The literature review showed that the use of LN-enabled micropayments as an incentive mechanism is a little-studied field. Building upon the concept of micropayments discussed earlier, the Lightning Network (LN) provides a scalable, low-cost platform to facilitate these transactions efficiently. This project aims to explore this further by collecting primary data with a questionnaire survey. The project is particularly interested in two groups: users of knowledge markets and those interested in blockchain and LN-integrated solutions. The questionnaire intends to understand whether users will be incentivised to ask and answer questions, respond faster, or put in more effort by varying levels of LN-based payment incentives. The results will inform payment incentive strategies for market applications and contribute to filling the gap in the literature below the previously studied levels.

An online questionnaire is the chosen distribution method since it requires less time investment than alternatives such as phone interviews. However, the risk of a low response rate was high. Therefore, the questions were closed and limited to 10 total, ensuring the total time to complete the questionnaire was reasonable. Google Forms was selected as the survey tool due to its user-friendliness and zero cost. It also provides multiple question types (e.g., multiple choice, grid) and can export data in real time to a .csv file or spreadsheet. Hence, the project could swiftly create pivot tables that update with each new submission, enabling near-instant analysis as responses came in.

By incorporating LN-enhanced micropayment models into Q&A platforms, the project seeks to determine if even minimal bounties (fractions of a penny) delivered via LN channels can alter user behaviour. Such an approach can test whether LN's efficiency and near-instant settlements enable practical incentives previously infeasible with conventional payment methods or solely on-chain solutions.

## 3.1 Questionnaire Structure

Consideration was given towards gender identification, responders not wanting to answer questions, and avoiding leading questions with biased phrasing. The introduction also clarifies that answers will be kept confidential. Figure 8 displays the most pertinent question relating to varying orders of magnitude of monetary incentive.

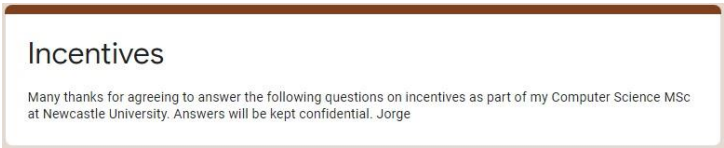

**Fig. 1**: Questionnaire introduction

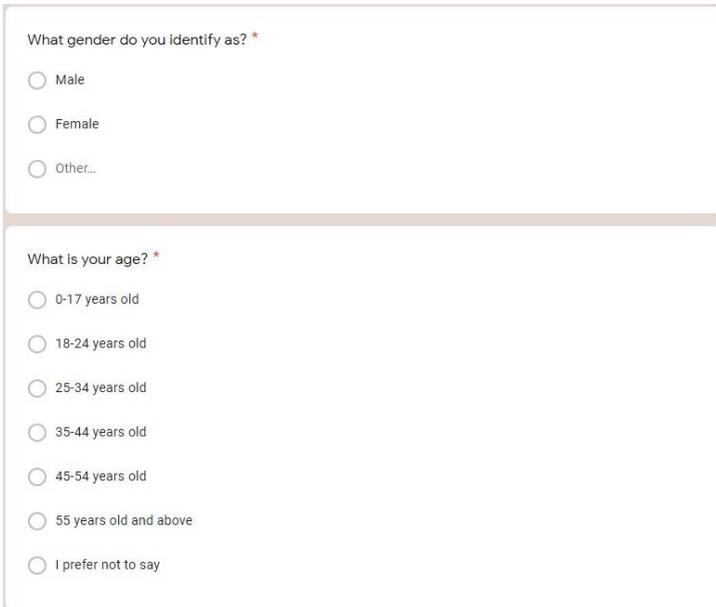

**Fig. 2**: Demographic questions 1 and 2

How often do you **ask** questions on knowledge sharing Q&A sites such as Stack Overflow, Yahoo Answers! and Quora? *

○ Several times a week

○ About once a week

○ Several times a month

○ About once a month

○ Less than once a month

○ Never

How often do you **answer** questions on knowledge sharing Q&A sites such as Stack Overflow, Yahoo Answers! and Quora? *

○ Several times a week

○ About once a week

○ Several times a month

○ About once a month

○ Less than once a month

○ Never

**Fig. 3**: Baseline Q&A questions 3 and 4

How many blockchain transactions have you (including businesses you own) participated in (sending or receiving) during the last 12 months? *

○ 0

○ 1-10

○ 11-100

○ 101-1000

○ Over 1000

**Fig. 4**: Baseline blockchain usage question 5

> A blockchain wallet integrated into a knowledge sharing Q&A site, where you could spend and earn money by asking and answering questions, would make you *
>
> ○ More likely to register
>
> ○ Neutral
>
> ○ Less likely to register

> When answering questions on a knowledge sharing Q&A site, you would rather receive *
>
> ○ Bulk payment at the end of each month
>
> ○ Micropayment for each answer in real time
>
> ○ I don't know

**Fig. 5**: Preference questions 6 and 7

> **Notes**
>
> For the following questions, you should assume that large payments and small micropayments will act as compensation in relation to the difficulty of the question and effort expended. For example, you might be able to answer 10 questions with a £0.20 bounty in the time it takes to answer one with a £2 bounty.
> Please indicate the extent to which you agree or disagree with the following statements.

**Fig. 6**: Scale question notes

The following mechanisms would incentivise you to **ask** more questions on knowledge sharing Q&A sites. *

|  | Strongly disagree | Disagree | Neutral | Agree | Strongly agree |
|---|---|---|---|---|---|
| Micropayments <= £1 | O | O | O | O | O |
| Payments > £1 | O | O | O | O | O |
| Monetary tips (user discretion, no limits) | O | O | O | O | O |
| Reputation points | O | O | O | O | O |

The following mechanisms would incentivise you to **answer** more questions on knowledge sharing Q&A sites. *

|  | Strongly disagree | Disagree | Neutral | Agree | Strongly agree |
|---|---|---|---|---|---|
| Micropayments <= £1 | O | O | O | O | O |
| Payments > £1 | O | O | O | O | O |
| Monetary tips (user discretion, no limits) | O | O | O | O | O |
| Reputation points | O | O | O | O | O |

**Fig. 7**: Q&A scale questions 8 and 9

**Fig. 8**: Payment scale question 10

1. What gender do you identify as?
2. What is your age?
3. How often do you **ask** questions on knowledge sharing Q&A sites such as Stack Overflow, Yahoo, Answers! and Quora?
4. How often do you **answer** questions on knowledge sharing Q&A sites such as Stack Overflow, Yahoo, Answers! and Quora?
5. How many blockchain transactions have you (including businesses you own) participated in (sending or receiving) during the last 12 months?
6. A blockchain wallet integrated into a knowledge sharing Q&A site, where you could spend and earn money by asking and answering questions, would make you... Investigate user opinion on blockchain wallet integration. Are most users positive? The result here directly maps to the project's value proposition since the project aim stipulates blockchain integration.
7. When answering questions on a knowledge-sharing Q&A site, you would rather receive... Investigate user opinion of micropayment schemes compared to the standard alternative. Would users prefer micropayments? Are they unsure? The result here directly maps to the project since its unique selling point compared to existing applications is the introduction of micropayment incentive models.

8. The following mechanisms would incentivise you to **ask** more questions on knowledge sharing Q&A sites… Examine whether micropayment incentives may increase the number of questions asked in consideration for the supposition that questions are priced out, compared against existing incentive models such as larger payments, tips, and reputation.
9. The following mechanisms would incentivise you to **answer** more questions on knowledge sharing Q&A sites… Explore the hypothesis in the project aims that micropayment incentives can improve the answer rate and time to respond to questions, compared against existing incentive models such as larger payments, tips, and reputation.
10. A question with the following bounties for an accepted answer would incentivise you to put in more effort to ensure your answer is accepted, compared to no bounty (£0)… Research payment incentive levels to determine trends and a point of significance, informing any LN-based micropayment strategies the project will undertake.

## 3.2 Questionnaire Results

The questionnaire study received a total of 32 responses from 22 males and 10 females. Responses were submitted anonymously after distribution amongst student and blockchain-focused Discord and Slack channels.

The overall aim of the study was to fill the gap in the literature for incentives below the $2 payment level, identified in the literature review.

The project investigated whether including a blockchain wallet into a Q&A application would incentivise users to register. A 3-point Likert scale was used, where 1 was less likely to register, 2 was neutral, and 3 was more likely to register.

| Question | Min | Max | Median | Std. Dev |
|---|---|---|---|---|
| **Blockchain Wallet Register Incentive** | 1 | 3 | 3 | 0.52 |

**Table 1**: Summary statistics of user likelihood to register for a blockchain integrated Q&A platform, revealing trends across varying blockchain engagement levels.

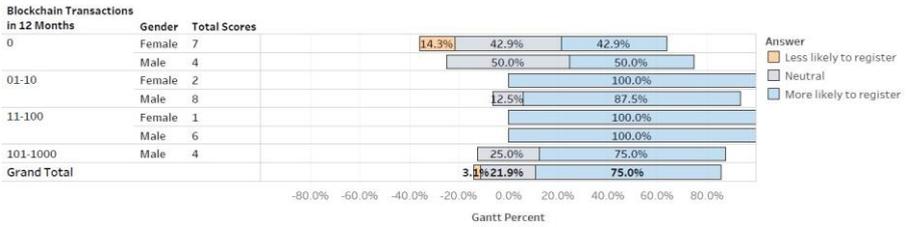

**Fig. 9**: Divergent stacked bar % blockchain wallet

The median score was 3, more likely to answer (see Table 1). As Figure 9 shows, 75% of respondents said they were more likely to register, with 21.9% remaining neutral and only 3.1% suggesting they would be less likely to register. Unsurprisingly, all of those less likely to register had participated in 0 blockchain transactions in the previous 12 months.

The project also sought to understand how different incentives might impact users' willingness to ask and answer questions. Further, the impact of varying question bounties between £0.001-£10 on users' effort expended to have an answer accepted was investigated.

A 5-point Likert scale was used, where 1 strongly disagree, 2 was disagree, 3 was neutral, 4 was agree, and 5 was strongly agree.

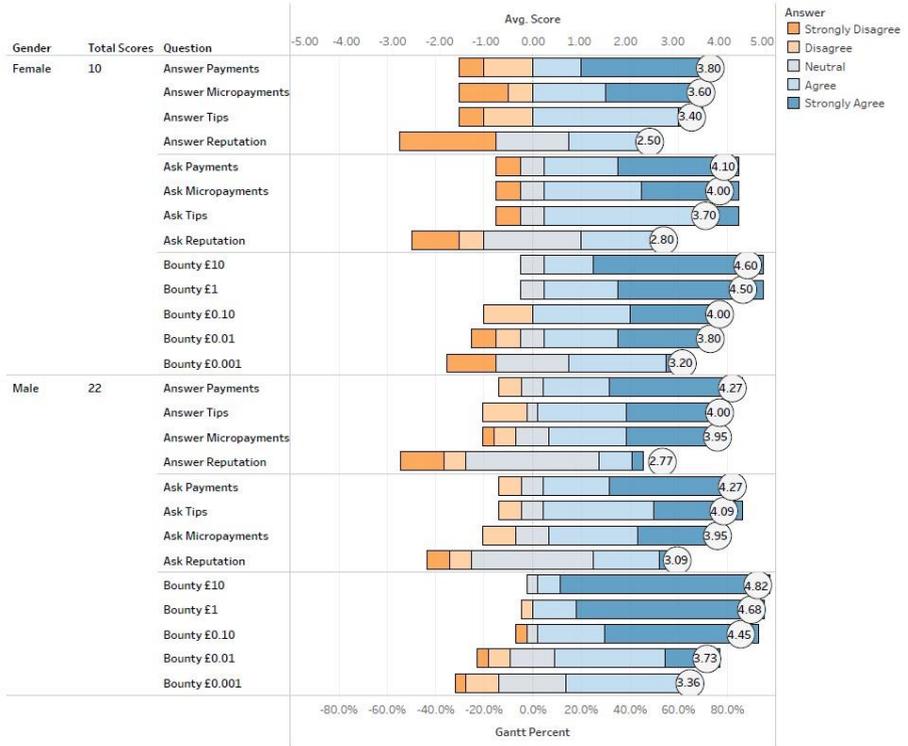

**Fig. 10**: Comparison of user incentives for asking and answering questions based on varying payment mechanisms, demonstrating a clear preference for financial incentives over reputation-based rewards.

| Question Group | Question | Min | Max | Median | Std. Dev |
|---|---|---|---|---|---|
| **Answer** | Answer Micropayments | 1 | 5 | 4 | 1.32 |
| | Answer Payments | 1 | 5 | 5 | 1.18 |
| | Answer Reputation | 1 | 5 | 3 | 1.15 |
| | Answer Tips | 1 | 5 | 4 | 1.18 |
| **Ask** | Ask Micropayments | 1 | 5 | 4 | 1.09 |
| | Ask Payments | 1 | 5 | 5 | 1.07 |
| | Ask Reputation | 1 | 5 | 3 | 1.02 |
| | Ask Tips | 1 | 5 | 4 | 0.97 |
| **Bounty** | Bounty £0.001 | 1 | 5 | 4 | 1.06 |
| | Bounty £0.01 | 1 | 5 | 4 | 1.16 |
| | Bounty £0.10 | 1 | 5 | 5 | 1.03 |
| | Bounty £1 | 2 | 5 | 5 | 0.71 |
| | Bounty £10 | 3 | 5 | 5 | 0.57 |

**Table 2**: 5-point Likert summary statistics

In Figure 10, the superimposed average score helps visualise the gulf between the top and bottom incentives. By looking at the distribution of answers across the Likert categories, we see that males and females are most incentivised to ask and answer questions by payments and least by a reputation scheme. Females generally prefer micropayments over tips, and males the opposite.

The average score implies agreement with payments, micropayments, and tips for answering questions across both genders, with a slight disagreement on the incentive of a reputation scheme. For asking questions, there is a more neutral opinion for reputation.

There is strong agreement at the £10 and £1 bounty levels, agreement at £0.10 and £0.01, and a fairly neutral opinion at the £0.001 level. This suggests that even as low as 1 penny might be enough motivation for users to put in significantly more effort into ensuring their answer is accepted.

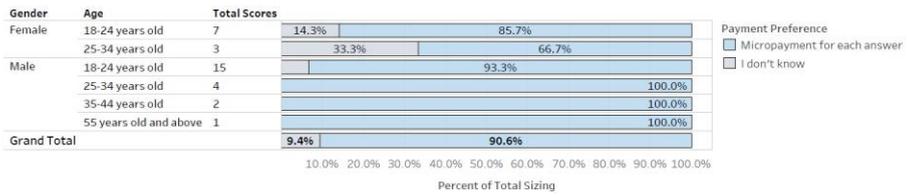

**Fig. 11**: Stacked bar % payment preference

Figure 11 shows there is a strong preference for a micropayment compensation strategy with 90.6% of respondents selecting it, nobody choosing bulk payment, and the remaining 9.4% being uncertain. Females and the 18-24 age bracket were the most uncertain. However, all men aged 25 and over preferred a micropayment strategy.

# 4  Conclusions and Future Directions

This study set out to investigate the potential of Lightning Network (LN)enabled micropayment strategies within the context of knowledge markets, specifically Q&A platforms. The results are encouraging: our evidence suggests that even extremely low-value incentives, as modest as £0.01, can positively influence user participation, effort, and the overall quality of contributions. By overcoming barriers associated with traditional payment processors and reducing transaction friction to near-negligible levels, the LN paves the way for a more fluid, user-centric economic model. In so doing, it supports a new era of the "Internet of Value" in which small, immediate financial incentives can meaningfully shape online behaviour.

While the results highlight LN's potential, they also underscore the necessity of addressing behavioural factors such as the perceived value of micropayments in varying cultural and economic contexts. This approach carries significant implications. Historically, knowledge sharing communities have been dominated by non-monetary incentives, such as reputation points and altruistic engagement. While these systems have produced thriving ecosystems, they also suffer from issues like unanswered questions, inconsistent answer quality, and user fatigue over time. Our findings hint that LN-based micropayments can complement and, in some cases, surpass existing incentive frameworks, making it easier for participants to receive tangible value for their contributions without large initial outlays. For instance, a niche programming Q&A forum could employ LN channels to encourage experts to address highly specific, lower-value questions, ultimately closing gaps in coverage and improving the platform's utility. Despite the promise, the study also reveals that questions remain. Users' self-reported intentions may not fully translate into consistent behaviour when actually spending their funds. Habit formation, market saturation, the perception of fairness, and the cognitive load of managing tiny payments still require deeper exploration. Additionally, while LN enables these micropayments technically, achieving long-term sustainability and widespread adoption involves grappling with regulatory, social, and infrastructural challenges. Thus, this work should be viewed as a starting point—an initial demonstration of feasibility rather than an exhaustive solution.

The findings open the door for extended analysis. Some users might gravitate to LN-based micropayment platforms to capitalise on their expertise quickly, while others may adopt a wait-and-see approach, preferring established Q&A communities. Future researchers could examine the interplay between these user segments, identifying the factors behind early adoption, continual usage, or migration away from micropayment-driven platforms.

Moreover, external market forces, such as cryptocurrency volatility, can also influence user uptake. Periods of high volatility might push users to hold rather than spend digital assets, whereas stable conditions may encourage frequent micropayment utilisation. Understanding these macroeconomic forces can help design more robust micropayment ecosystems.

## 4.1 Future Directions

To strengthen and broaden the understanding of LN-driven micropayment economies, several future research directions are proposed:

1. **Longitudinal, Real-World Implementations:** Future work should move beyond controlled surveys and implement LN-based micropayment mechanisms directly into live Q&A platforms or experimental prototypes. Monitoring user behaviour over extended periods—months or even years—will reveal whether initial enthusiasm for micropayments wanes or stabilises. Such longitudinal field studies might involve gradually introducing micropayments to a subset of users, adjusting bounty sizes dynamically, or introducing time-based incentives (e.g., higher bounties for quicker responses) to see if these factors yield sustained improvements in participation and answer quality.
2. **Comparative Studies Across Different Knowledge Domains:** Different fields of expertise may respond differently to micropayment incentives. Highly technical communities, such as advanced coding forums, might place a higher value on even minuscule bounties if they acknowledge the complexity of questions being asked. Conversely, more general interest Q&A sites could require larger incentives or the combination of micropayments with reputation metrics to drive engagement. Comparative studies can test whether LN-driven models are universally applicable or must be tailored to the cultural norms, complexity, and difficulty gradients within different domains.
3. **Hybrid Incentive Systems with Gamification:** Although micropayments are promising, hybrid incentive systems combining LN-based financial rewards with non-monetary motivators (e.g., digital badges, leaderboards, time-limited challenges) warrant exploration. A system could, for example, provide small LN-denominated rewards for each accepted answer while also awarding symbolic badges for consistent participation or innovative solutions. Understanding the synergy between monetary and non-monetary incentives can help platforms maintain user engagement, reduce attrition, and foster community spirit while still leveraging micropayments to ensure quality and timeliness.
4. **Integration of Smart Contracts and Programmable Logic:** Future research can delve into more sophisticated payment schemes facilitated by LN and smart contracts. For example, an escrow-based model could lock a bounty within a smart contract that only releases funds upon meeting predefined criteria—such as community-voted acceptance, achieving a certain score from an automated quality checker, or passing a verification step by trusted moderators. This could ensure that even tiny payments maintain integrity and discourage low-effort spam. Similarly, tiered, dynamic, or subscription-based payment models could be introduced, allowing experts to establish LN channels with power users who pay incrementally for premium assistance.
5. **Impact on User Privacy and Data Monetisation Models:** LN-based micropayment markets have the potential to reduce reliance on surveillance capitalism, as revenue streams shift from advertising and data harvesting to direct user payments. Future studies could quantify the reduction in data exploitation when users are able to "pay their way" out of personalised ads. Researchers could measure user satisfaction, privacy sentiment, and perceived control over personal data in LN-enabled environments. Additionally, exploring stablecoins or other value-pegged assets on LN may mitigate volatility concerns, making it more appealing for users hesitant to transact in volatile cryptocurrencies.
6. **Regulatory and Compliance Considerations:** As micropayments become more common, regulators will need guidance on how to treat these transactions. Research can focus on designing LN-based systems that comply with AML (Anti-Money Laundering) and KYC (Know Your Customer) regulations without sacrificing user experience. Striking the right balance between financial compliance and low-friction user onboarding remains an open question. Studies that engage policymakers, legal experts, and industry stakeholders could propose frameworks that ensure consumer protection, fraud prevention, and tax compliance for micropayment ecosystems.

7. **Scalability, Liquidity, and Network Health:** The LN relies on efficient routing and adequate channel liquidity. Future work could simulate large-scale LN-based Q&A platforms to understand the impact on liquidity requirements and routing fees. As the volume of microtransactions grows, do network conditions remain favourable, or will new optimisation strategies be required? Research into automated channel management, dynamic fee adjustments, and liquidity pools specifically tailored for Q&A or content marketplaces could ensure that LN scaling keeps pace with user demand.

8. **User Education, Onboarding, and Interface Design:** Micropayment adoption depends on user-friendly interfaces and accessible educational resources. Future studies can investigate how best to present LN wallets, explain the concept of micropayments, and guide newcomers through their first low-value transactions. Trials might compare different UI layouts, tutorial styles, or incentives for first-time users. Reducing cognitive load and ensuring that transacting in LN-based micropayments feels as natural as clicking a "like" button will be critical for mass adoption.

9. **Cross-Platform and Interoperability Initiatives:** Knowledge markets do not exist in isolation. Future research could explore cross-platform interoperability, where LN-enabled Q&As seamlessly interact with other LNintegrated services—such as paid content aggregators, microtask platforms, or data marketplaces—forming a larger micro-economy. Understanding how users move between these interconnected ecosystems, how their reputation or creditworthiness transfers, and how pricing equilibria emerge across different domains could provide a roadmap for a holistic Internet of Value that transcends traditional platform boundaries.

10. **Cultural and Geographical Factors:** Micropayment acceptance may vary widely by region due to differences in economic conditions, technology adoption rates, cultural attitudes toward money, and existing financial infrastructures. Comparative international studies could reveal whether LNbased micropayment models need tailoring to local conditions. For example, lower-income regions may appreciate and utilise micropayments differently than wealthier markets, potentially discovering that micropayments serve as a powerful tool for economic empowerment and skill monetisation in communities historically underserved by global financial systems.

Another critical aspect involves user education on blockchain technology and LN-based payments. To maximise adoption, future studies should explore effective methods for simplifying user onboarding, including intuitive wallet interfaces and step-by-step tutorials tailored to diverse demographic groups.

Ultimately, LN-driven micropayments represent a paradigm shift in how value is exchanged online. By breaking the economic and infrastructural constraints that traditionally impeded small transactions, LN opens the door to a world where knowledge sharing is fluid, economically sustainable, and less dependent on monolithic platforms or advertising models. With continued research, experimentation, and refinement in the areas outlined above, the vision of a more equitable, user-driven, and privacy-respecting digital economy can move closer to reality. This study is but one step in that direction, and the future landscape of LN-integrated micropayment ecosystems promises a rich tapestry of challenges, opportunities, and transformative possibilities.